\documentclass[11pt]{article}
\usepackage{amssymb,amsmath,amsfonts}
\usepackage{graphicx}
\usepackage{graphics}
\usepackage{dcolumn}
\usepackage{eepic,epsfig}

\begin{document}

\title{High power Cherenkov radiation from a relativistic particle rotating
around a dielectric ball }
\author{L.Sh. Grigoryan\thanks{%
E-mail address: levonshg@web.am}, H.F. Khachatryan, S.R. Arzumanyan\thanks{%
E-mail address: arzumsr@yandex.ru}, M.L. Grigoryan \\
\textit{Institute of Applied Problems in Physics}\\
\textit{25 Hr. Nersessian Str., 375014 Yerevan, Armenia}\\
}
\date{\today }
\maketitle

\begin{abstract}
Some characteristic features in the radiation from a relativistic electron
uniformly rotating along an equatorial orbit around a dielectric ball have
been studied. It was shown that at some harmonics, in case of weak
absorption of radiation in the ball material, the electron may generate
radiation field quanta exceeding in several dozens of times those generated
by electron rotating in a continuous, infinite and transparent medium having
the same real part of permittivity as the ball material. The rise of high
power radiation is due to the fact that electromagnetic oscillations of
Cherenkov radiation induced along the trajectory of particle are partially
locked inside the ball and superimposed in nondestructive way.
\end{abstract}



\bigskip

\section{Introduction}

\label{section1}

Due to such unique properties as the high intensity, high degree of
collimation, and wide spectral range (see \cite{1}-\cite{6} and references
therein) Synchrotron Radiation (SR) serves as an extraordinary research tool
for advanced studies in both the fundamental and applied sciences. Nowadays
it is used worldwide by thousands of scientists in many disciplines like
physics, chemistry, material science and structural biology. These
applications motivate the importance of analyzing various mechanisms for
control of SR parameters. From this point of view it is of interest to study
the influence of medium on spectral and angular distributions of SR.
Investigations of this kind are also important for a number of astrophysical
problems \cite{4,7}.

The characteristics of high-energy electromagnetic processes in the presence
of material are essentially changed by giving rise to new types of
phenomena, the well-known examples of which are Cherenkov \cite{8}-\cite{10}
and diffraction radiation. The operation of many devices intended for
production of electromagnetic radiation is based on interactions of
relativistic electrons with matter (see, e.g., \cite{11}).

As was shown in \cite{12}, where SR from a charged particle rotating in a
homogeneous medium was considered (see also \cite{10,13}), the interference
between SR and Cherenkov Radiation (CR) had interesting consequences. New
interesting phenomena occur in case of inhomogeneous media. A well-known
example here is the transition radiation. In particular, the interfaces
between media can be used for monitoring the flows of radiation emitted in
various systems. In a series of papers initiated in \cite{14}-\cite{16} it
was shown that the interference between SR and CR induced at boundaries of
spherical or cylindrical configuration leads to interesting effects.

Investigations of radiation from a charge rotating along an equatorial orbit
about/inside a dielectric ball showed \cite{17,18} that when the Cherenkov
condition for the ball material and particle speed is satisfied, there
appear high narrow peaks in the spectral distribution of the number of
quanta emitted to outer space at some specific values of the ratio of
ball-to-particle orbit radii. In the vicinity of these peaks the radiated
energy exceeds the corresponding value for the case of homogeneous and
unbounded medium by several orders of magnitude. However in \cite{17,18} the
phenomena of absorption and dispersion of electromagnetic waves inside the
ball material were not taken into account. The allowance for these phenomena
was made in \cite{Mher}. In the present paper a new characteristic feature
of such a high power radiation was revealed and a simple model that
disclosed the physical cause of its generation was suggested.

It is worthwhile to point out that a similar phenomenon (less pronounced)
takes place in the cylindrical symmetry case \cite{19}-\cite{25}. E.g., the
radiation emitted (i) from a longitudinal charged oscillator moving with
constant drift velocity along the cylinder axis, and (ii) from a charged
particle moving along a circle around a dielectric cylinder or (iii)\ along
a helical orbit inside the cylinder are investigated in \cite{19}-\cite{21},
\cite{22} and \cite{23}-\cite{25} respectively. The latter type of motion is
used in helical undulators for generating electromagnetic radiation in a
narrow spectral interval. It is shown that under similar Cherenkov condition
for permittivity of cylinder and the particle speed, high narrow peaks are
present in the spectral-angular distribution for the number of radiated
quanta. In the vicinity of these peaks the radiated energy exceeds the
corresponding value for homogeneous medium case by several orders of
magnitude.

The content of paper is organized as follows: in Section \ref{section2}\ the
description of problem is given, and in Section \ref{section3} the method of
solution is described and the final analytical expression for the intensity
of radiation from a relativistic particle rotating about a ball is given.
Numerical results are presented in Section \ref{section4}. The
characteristic features of phenomenon are discussed in Section \ref{section5}%
. The cause of radiation amplification is established in Section \ref%
{section6}. Similar processes of amplification typical for a chain of
equidistant particles/bunches, are considered in the next section. In the
last section the main results of paper are summarized.

\section{The description of problem}

\label{section2}

\begin{figure}[tbph]
\begin{center}
\epsfig{figure=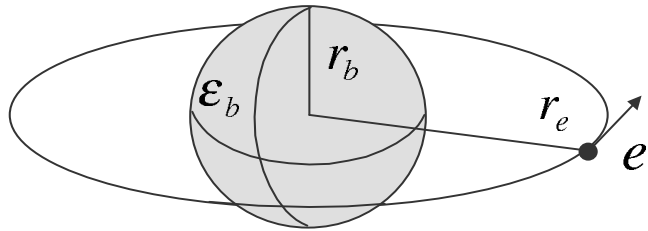,width=8.5cm,height=3cm}
\end{center}
\caption{A relativistic electron rotating about a dielectric ball in its
equatorial plane. }
\label{fig1}
\end{figure}

Now consider the uniform rotation of relativistic electron in the equatorial
plane of a dielectric ball in the magnetic field in empty space (see Fig. %
\ref{fig1}).

The permittivity of medium is the following step function of radial
coordinate r:%
\begin{equation}
\varepsilon \left( r\right) =\varepsilon _{b}+\left( 1-\varepsilon
_{b}\right) \Theta \left( r-r_{b}\right) ,  \label{epsr}
\end{equation}%
where $r_{b}$ is the radius of ball, and $\varepsilon _{b}=\varepsilon
_{b}^{^{\prime }}+i\varepsilon _{b}^{^{\prime \prime }}$ is the complex
valued permittivity of the ball material. We assume that the braking of
electron due to the emission of radiation is compensated by an external
influence (e.g., the electric force) that compels the particle to turn
uniformly in a circle.

It is convenient to introduce the following dimensionless quantity:$\qquad
\qquad \qquad $%
\begin{equation}
\text{w}_{kT}/\hbar \omega _{k}\equiv n_{k}  \label{nk}
\end{equation}%
(the number of emitted quanta). Here w$_{kT}$\ is the energy radiated at
frequency $\omega _{k}=k\omega _{0}$ during one period $T=2\pi /\omega _{0}$
of electron gyration, $k=1;2;3...$ is the harmonic number, and $\hbar \omega
_{k}$\ is the energy of corresponding electromagnetic wave quantum.

It is known \cite{10} that if the space as a whole is filled with a
transparent substance (with constant $\varepsilon $) then$\qquad \qquad
\qquad \qquad $%
\begin{equation}
n_{k}\left( \infty ;\text{v},\varepsilon \right) =\dfrac{n_{0}}{\beta \sqrt{%
\varepsilon }}\left[ 2\beta ^{2}J_{2k}^{^{\prime }}\left( 2k\beta \right)
+\left( \beta ^{2}-1\right) \int_{0}^{2k\beta }J_{2k}\left( x\right) dx%
\right] ,  \label{nkinfinity}
\end{equation}%
where $n_{0}=2\pi e^{2}/\hbar c\cong 0.0459$\ , $\beta =$v$\sqrt{\varepsilon
}/c$\ and $J_{k}\left( x\right) $\ is the Bessel function of integer order.
The case $\varepsilon =1$ of this formula corresponds to the synchrotron
radiation in vacuum (see, e.g., \cite{2,3}).

We aimed at:

\begin{itemize}
    \item calculation of $n_{k}=n_{k}\left( \text{ball}\right) $, when $%
\varepsilon \left( r\right) $\ is the step function (\ref{epsr})
and

    \item demonstration that in case of weak absorption of radiation in the
ball material and at special choice of the values of $r_{b}$\ and $%
\varepsilon _{b}$\ , the number of emitted quanta
$$ n_{k}\left( \text{ball};\text{v}%
,r_{b}/r_{e},\varepsilon _{b}\right) \gg n_{k}\left( \infty ;\text{v}%
,\varepsilon _{b}^{^{\prime }}\right) . $$
\end{itemize}

\section{The final formula}

\label{section3}

We have obtained \cite{14,15,17} (see also \cite{18}) the following
expression\qquad\ \qquad \qquad
\begin{equation}
n_{k}\left( \text{ball;v},x,\varepsilon _{b}\right) =\dfrac{2n_{0}}{k}%
\sum_{s=0}^{\infty }\left( \left\vert a_{kE}\left( s\right) \right\vert
^{2}+\left\vert a_{kH}\left( s\right) \right\vert ^{2}\right)  \label{nkball}
\end{equation}%
for the number of quanta emitted by the electron during one revolution. Here
\ $x\equiv r_{b}/r_{e}<1$ and
\begin{eqnarray}
a_{kE}\left( s\right) &=&kb_{l}\left( E\right) P_{l}^{k}\left( 0\right)
\sqrt{\left( l-k\right) !/l\left( l+1\right) \left( 2l+1\right) \left(
l+k\right) !},\quad l=k+2s,  \notag \\
a_{kH}\left( s\right) &=&b_{l}\left. \left( H\right) \sqrt{\dfrac{\left(
2l+1\right) \left( l-k\right) !}{l\left( l+1\right) \left( l+k\right) !}}%
\dfrac{dP_{l}^{k}\left( y\right) }{dy}\right\vert _{y=0},\quad l=k+2s+1
\label{akE,akH}
\end{eqnarray}%
are dimensionless amplitudes describing the contributions of electrical (%
\textit{E}) and magnetic (\textit{H}) type multipoles respectively, $%
P_{l}^{k}\left( y\right) $ are the associated Legendre polynomials, and $%
b_{l}\left( E\right) $, $b_{l}\left( H\right) $ are the following factors
depending on $k,$v$,x$ and $\varepsilon _{b}$:
\begin{eqnarray}
b_{l}\left( H\right) &=&iu\left[ j_{l}\left( u\right) -h_{l}\left( u\right)
\dfrac{\left\{ j_{l}\left( xu_{b}\right) ,j_{l}\left( xu\right) \right\} _{x}%
}{\left\{ j_{l}\left( xu_{b}\right) ,h_{l}\left( xu\right) \right\} _{x}}%
\right] ,\quad u=k\text{v}/c,\quad u_{b}=k\text{v}\sqrt{\varepsilon _{b}}/c,
\notag \\
b_{l}\left( E\right) &=&\left( l+1\right) b_{l-1}\left( H\right)
-lb_{l+1}\left( H\right) +x^{-2}\left( 1-\varepsilon _{b}\right) \left[ j_{%
\underline{l-1}}\left( xu_{b}\right) +j_{\underline{l+1}}\left(
xu_{b}\right) \right] \times  \notag \\
&&\times \left[ h_{\underline{l-1}}\left( u\right) +h_{\underline{l+1}%
}\left( u\right) \right] \frac{l\left( l+1\right) u_{b}j_{l}\left(
xu_{b}\right) }{lz_{l-1}^{l}+\left( l+1\right) z_{l+1}^{l}}.  \label{blH,blE}
\end{eqnarray}%
Here $h_{l}\left( y\right) =j_{l}\left( y\right) +i$n$_{l}\left( y\right) $,
and $j_{l}\left( y\right) $, \ n$_{l}\left( y\right) $ are spherical Bessel
and Neumann functions respectively. In (\ref{blH,blE}) we used the following
notations:
\begin{eqnarray}
\left\{ a\left( x\alpha \right) ,b\left( x\beta \right) \right\} _{x}
&\equiv &a\dfrac{\partial b}{\partial x}-\dfrac{\partial a}{\partial x}b,
\notag \\
f_{\underline{l}}\left( x\right) &\equiv &f_{l}\left( x\right) /\left\{
j_{l}\left( xu_{b}\right) ,h_{l}\left( xu\right) \right\} _{x},
\label{telnju} \\
z_{\nu }^{l} &\equiv &\dfrac{uj_{\nu }\left( xu_{b}\right) h_{l}\left(
xu\right) \varepsilon _{b}-u_{b}j_{l}\left( xu_{b}\right) h_{\nu }\left(
xu\right) }{uj_{\nu }\left( xu_{b}\right) h_{l}\left( xu\right)
-u_{b}j_{l}\left( xu_{b}\right) h_{\nu }\left( xu\right) }.  \notag
\end{eqnarray}%
\qquad \qquad

In the absence of dielectric ball $\left( \varepsilon _{b}=1\right) $ the
calculations by means of our formula (\ref{nkball}) give the same results as
those obtained using the well-known synchrotron radiation theory formula (%
\ref{nkinfinity}).

\section{ The results of numerical calculations}

\label{section4}

Now consider the radiation generated by the electron at some harmonic $%
\omega _{k}=k\omega _{0}$, e.g., for $k=8$.
\begin{figure}[tbph]
\begin{center}
\epsfig{figure=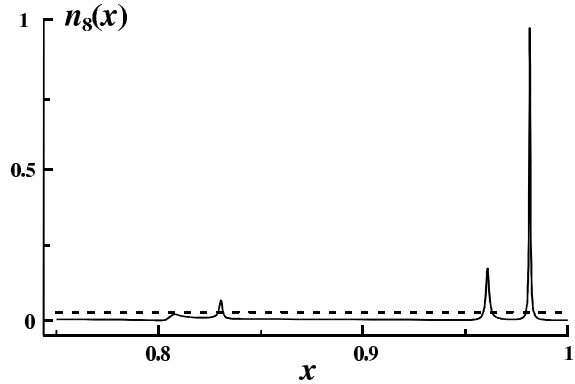,width=7.5cm,height=5cm}
\end{center}
\caption{The number $n_{k}(x)$ of electromagnetic field quanta generated per
revolution of electron about a dielectric ball depending on the ratio $%
x=r_{b}/r_{e}$ of the radius of ball and that of electron orbit. The
harmonic number $k=8$, the electron energy $E_{e}=2MeV$, the radius of its
orbit $r_{e}=3.69cm$. The corresponding radiation is at the wavelength $3cm$%
(in vacuum). The ball is made of melted quartz. The dielectric losses of
energy inside the ball material are taken into account. For explanation see
the text.}
\label{fig2}
\end{figure}
In Fig.\ref{fig2} the number $n_{8}\left( x\right) $\ of emitted quanta was
plotted versus the ratio $x=r_{b}/r_{e}$\ of ball to electron orbit radii.
The energy and the orbit radius of the particle were taken to be $E_{e}=2MeV$%
\ and \ \ $r_{e}=3.69cm$ respectively. Here corresponding to the 8th
harmonic is the radiation at frequency $\omega _{8}/2\pi =10^{10}Hz$\ and
wavelength $\lambda _{8}=3cm$ (in vacuum). The number of quanta was
calculated in \cite{Mher} by formula (\ref{nkball}) with due regard for the
dispersion and dielectric losses of energy inside the substance of ball
under the assumption that the ball is made of melted quartz, the value of
permittivity of which is \cite{26,27}:
\begin{equation}
\varepsilon _{b}\left( \omega _{8}\right) =\varepsilon _{b}^{^{\prime
}}+i\varepsilon _{b}^{^{\prime \prime }}=3.78\left( 1+0.0001i\right)
\label{epsom8}
\end{equation}%

According to (\ref{nkinfinity}) the number of quanta emitted on the 8th
harmonic by electron with above values of $E_{e}$\ and $r_{e}$\ at rotation
in a continuous, infinite and transparent medium with $\varepsilon =3.78$\
would be \cite{10}
\begin{equation}
n_{8}\left( \infty \right) \cong 0.0274  \label{nkbesk}
\end{equation}%
(the horizontal dashed line in Fig.\ref{fig2}). If the particle did not
rotate, but moved rectilinearly in the same medium with energy $E_{e}=2MeV$,
then during the period of time $T=2\pi /\omega _{0}$\ it would have emitted
in a narrow frequency band $\Delta \omega =\omega _{0}$
\begin{equation}
n_{\Delta \omega }\left( \infty \right) =\left( \text{v}/c-c/\text{v}%
\varepsilon \right) n_{0}\cong 0.0318  \label{nkdeltainfinity}
\end{equation}%
quanta (see, e.g., \cite{8}-\cite{10}). It is evident that $n_{\Delta \omega
}\left( \infty \right) \sim n_{8}\left( \infty \right) $. In the absence of
ball (synchrotron radiation \cite{1}-\cite{6})
\begin{equation}
n_{8}\left( \text{vac}\right) \cong 0.00475.  \label{n8vac}
\end{equation}

According to Fig.\ref{fig2}, $n_{8}\left( x\right) \sim n_{8}\left( \text{vac%
}\right) $\ practically for all $x$ except for $0.8<x<0.85$\ and $0.95<x<1$.
There are peaks in these ranges and for the highest one

\begin{equation}
x^{\ast }=0.9815\quad \text{and\quad }n_{8}\left( \text{ball};x^{\ast
}\right) \cong 0.951\quad \Longrightarrow \quad n_{8}\left( \text{ball}%
;x^{\ast }\right) /n_{8}\left( \infty \right) \cong 35.
\label{n8ball/n8infinity}
\end{equation}

Here more than 30-fold magnification of $n_{8}$\ (from \ $\approx 0.03$ to $%
\approx 1.0$) is due to the forced uniform rotation of electron around the
ball under the action of an external energy source. The corresponding value
of ball radius $r_{b}=3.62cm$\ and, consequently, the distance $r_{e}-r_{b}$%
\ between the rotating electron and the ball surface should be $0.7mm$.
Apparently,
\begin{equation}
n_{8}\left( \text{vac}\right) <<n_{\Delta \omega }\left( \infty \right) \sim
n_{8}\left( \infty \right) <<n_{8}\left( \text{ball};x^{\ast }\right) .
\label{n8vac-nball}
\end{equation}

Similar results may be obtained for a series of other values of $k>>1$, as
well as for electrons with $1\leq E_{e}\leq 5MeV$\ energy and balls with $%
1\leq \varepsilon _{b}^{^{\prime }}\leq 5$ and $\varepsilon _{b}^{^{\prime
\prime }}/\varepsilon _{b}^{^{\prime }}<<1$.

\section{Peculiarities of the phenomenon}

\label{section5}

It is significant that the condition
\begin{equation}
n_{k}\left( \text{ball};\text{v},x^{\ast },\varepsilon _{b}\right)
>>n_{k}\left( \infty ;\text{v},\varepsilon _{b}^{^{\prime }}\right)
\label{nkball-nkinfinity}
\end{equation}%
holds

(A) only for a number of values of $r_{b}/r_{e}=x^{\ast }$\ and also

(B) when simultaneously the Cherenkov condition
\begin{equation}
\text{v}_{\ast }\sqrt{\varepsilon _{b}^{^{\prime }}}/c>1  \label{Cherenkov}
\end{equation}%
for speed v$_{\ast }\equiv x^{\ast }$v$=r_{b}\omega _{0}$\ of particle
"image" on the ball surface (see. Fig.\ref{fig3}) and the real part $%
\varepsilon _{b}^{^{\prime }}$ of the permittivity of ball material is
satisfied.

For all peaks (\ref{nkball-nkinfinity}) the characteristic propagation
lengths $l=2c\sqrt{\varepsilon _{b}^{^{\prime }}}/\omega _{k}\varepsilon
_{b}^{^{\prime \prime }}$, after passage of which the amplitude of
electromagnetic oscillations decreases in $e$ times, are always many times
greater than the ball radius:\qquad
\begin{equation}
l>>r_{b}.  \label{l>>rb}
\end{equation}%
For this reason the third condition may be formulated as:

(C) inside the ball the electromagnetic oscillations may pass the distances
that exceed many times the radius $r_{b}$\ of ball (before oscillations are
absorbed or leave the boundaries of ball).

\section{The reason of radiation amplification}

\label{section6}

\bigskip If the electron is relativistic, then in addition to the
synchrotron radiation it may also generate Cherenkov radiation. Its rise is
attributed to the fact that the field coupled with the relativistic electron
partially penetrates the ball and rotates together with the particle. At
small distances from the surface of ball: $r_{e}\approx r_{b}$, the speed of
this field displacement inside the ball may be larger than the phase speed $%
c/\sqrt{\varepsilon _{b}^{^{\prime }}}$\ of light in the ball material, and
then CR with wavelengths \cite{8}-\cite{10}
\begin{equation}
\lambda _{k}/\sqrt{\varepsilon _{b}^{^{\prime }}}\geq \left(
r_{e}-r_{b}\right) .  \label{wavelength}
\end{equation}%
is generated inside the ball. In this section for visual explanation of
numerical results given in Sections \ref{section4},\ref{section5} we shall
avail of the following simplified model.

\begin{figure}[tbph]
\begin{center}
\epsfig{figure=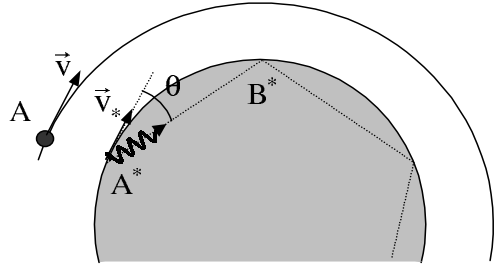,width=6.5cm,height=3.5cm}
\end{center}
\caption{A rotating electron in point $A$ along with the synchrotron
radiation generates a pulse of Cherenkov radiation (the wave line) in the
vicinity of point $A^{\ast }$\ inside the ball (the speed of electron field
displacement inside the ball exceeds the phase speed of light in the ball
material). The pulse travels to the vicinity of point \ $B^{\ast }$and at
least some part of radiation is subjected to an "almost total" internal
reflection at $\protect\theta $ angle.}
\label{fig3}
\end{figure}

Let us assume that at the moment $t_{A}$\ the electron is in point $A$ of
orbit and its image is in point \ $A^{\ast }$ (see. Fig.\ref{fig3}).
\begin{figure}[tbph]
\begin{center}
\epsfig{figure=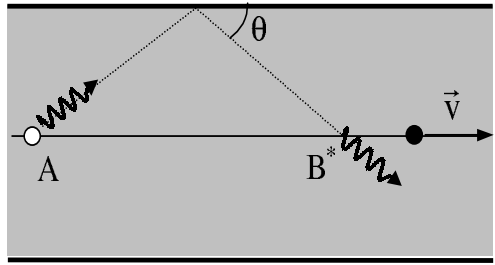,width=6.5cm,height=3.5cm}
\end{center}
\caption{A relativistic charged particle moves along the axis of circular
section waveguide filled with dielectric. The particle generates a pulse of
Cherenkov radiation in the vicinity of point $A$ (the wave line). At its
propagation the pulse at some instant crosses the trajectory of its
faster-than-light source (axis of waveguide) in the vicinity of point $%
B^{\ast }$\ at the same angle $\protect\theta $, at which it was emitted. At
that moment the pulse will fall behind the particle that will be found more
to the right at some point $B$.}
\label{fig4}
\end{figure}
At this instant the electron generates CR along with the synchrotron
radiation, or to be more exact, generates CR pulse $W\left( A^{\ast };t\geq
t_{A}\right) $ (the wavy line) in the vicinity of point $A^{\ast }$\ inside
the ball near the surface. The direction of CR is characterized by some
radiation angle $\theta $. We shall use the following ansatz that $\theta $\
is determined by the Cherenkov condition%
\begin{equation}
\cos \theta =c/\text{v}_{\ast }\sqrt{\varepsilon _{b}^{^{\prime }}}
\label{costet}
\end{equation}%
by analogy with the case of faster-than-light particle propagating
rectilinearly in a continuous and infinite dielectric medium \cite{8}-\cite%
{10}. In (\ref{costet}) v$_{\ast }=$v$r_{b}/r_{e}$\ is the speed of particle
"image" on the surface of ball. According to such a simplified approach, the
condition B of Section \ref{section5} establishes that the electron
generates CR inside the dielectric ball. The pulse $W\left( A^{\ast };t\geq
t_{A}\right) $\ travels the distance $A^{\ast }B^{\ast }$\ to the vicinity
of point $B^{\ast }$ by the surface of ball and is partially absorbed by the
ball material. In (\ref{costet})
\begin{equation*}
\cos \theta >1/\sqrt{\varepsilon _{b}^{^{\prime }}}
\end{equation*}%
and, consequently, at least some part of radiation with the wavelength
\begin{equation}
\lambda _{k}/\sqrt{\varepsilon _{b}^{^{\prime }}}<<r_{b}  \label{lamrb}
\end{equation}%
(according to the laws of geometrical optics) will be subjected to an
"almost total" internal reflection at $\theta $\ angle. This process will
continue till $W\left( A^{\ast };t\geq t_{A}\right) $ is totally absorbed
and/or leaves the ball. At $\varepsilon _{b}^{^{\prime }}/\varepsilon
_{b}^{^{\prime \prime }}>>1$\ (small adsorption) the pulse $W\left( A^{\ast
};t\geq t_{A}\right) $\ "will have time" to reflect from the internal
surface of ball
\begin{equation}
m\approx \dfrac{l}{A^{\ast }B^{\ast }}=\dfrac{\varepsilon _{b}^{^{\prime
}}/\varepsilon _{b}^{^{\prime \prime }}}{k\tan \theta }>>1
\label{reflectiontimes}
\end{equation}%
times and, consequently, to travel the distance $mA^{\ast }B^{\ast }\sim
mr_{b}>>r_{b}$ inside the ball in agreement with condition C of the previous
section. Over the period of time $T_{l}=l\sqrt{\varepsilon _{b}^{^{\prime }}}%
/c$, during which the oscillation amplitude in the pulse $W\left( A^{\ast
};t\geq t_{A}\right) $ decreases in $e$ times, the electron will make
\begin{equation}
N\equiv \left[ T_{l}/T\right] =\left[ \varepsilon _{b}^{^{\prime }}/\pi
k\varepsilon _{b}^{^{\prime \prime }}\right] >>1
\label{Number of revolution}
\end{equation}%
revolutions along the orbit ($\left[ \tau \right] $means the integer part of
number $\tau $). In this process the electron will appear in the vicinity of
point $A$ many times at instants $t_{A}+T$; $t_{A}+2T$; \ldots\ $t_{A}+NT$,
and its image will appear in the vicinity of point $A^{\ast }$\ as much
times again and each time will generate CR pulse. All \ $N+1$ pulses $%
W\left( A^{\ast };t\geq t_{A}+sT\right) $\ with $s=0;1;2...N$ \ will
propagate inside the confined space (ball) along identical "trajectories".
For the highest peak in Fig.\ref{fig2} $m\approx 806$\ and $N=398.$

The aforesaid is applicable to each point along the electron orbit. Being
superimposed in the confined space (ball), the electromagnetic oscillations
of numerous CR pulses will, naturally, damp each other. In this case the
influence of ball will be small and the total radiation will not differ
strongly from the synchrotron radiation. This is precisely the case for the
most values of $x$ (see Fig.\ref{fig2}). However, in some special cases the
deviations from such a picture are possible. These take place in ranges $%
0.8<x<0.85$\ and $0.95<x<1$ shown in Fig.\ref{fig2}. Inside these ranges at
some special choice of $x=r_{b}/r_{e}\equiv x^{\ast }$(cf. with condition A
in Section \ref{section5}) the superposition of electromagnetic oscillations
will be close to the in-phase one. In this case the total field may prove
considerably more strong than that at $x\neq x^{\ast }$. The force that
brakes the gyration of electron will also increase and \textit{the extra
work of external force compelling the electron to rotate uniformly along the
orbit will be spent for generation of more powerful CR. }This is precisely
the case at $x^{\ast }=0.9815$\ in Fig.\ref{fig2}.

In Section \ref{section4} we determined the values of $x=x^{\ast }$\ from
the requirement of the presence of peak, $n_{k}\left( \text{ball};x^{\ast
}\right) >>n_{k}\left( \infty \right) $, by numerical calculations by
formula (\ref{nkball}), disregarding (\ref{costet}) and neglecting the laws
of geometrical optics.

\begin{enumerate}
    \item Thus, in general, the influence of ball on
radiation from the particle is small and the total radiation
differs slightly from synchrotron radiation.
    \item However, at some special values of the ratio $r_{b}/r_{e}\equiv x^{\ast }$
high power CR will be generated.
    \item Its rise is due to the fact that the electromagnetic oscillations of CR
induced by the particle\ along all the trajectory are partially
locked inside the ball and are superimposed in nondestructive way.

\end{enumerate}

\section{Similar processes of amplification}

\label{section7}

\subsection{Amplification of transition radiation [28,29]}

\label{section7.1}

In 1996 and 1997 Prof. Wiedemann with co-workers experimentally observed a
similar phenomenon for Transition Radiation (TR) from a chain of
relativistic electron bunches of sub-picosecond duration from the Stanford
University Short Intense Electron source (SUNSHINE).

At the output of TR radiator (metallic foil) the radiation from one electron
bunch was directed with the help of a special system of mirrors (influence
of boundaries) back to the radiator so, that the radiation pulse arrived
there at the instant of incidence of the other electron bunch to generate
new TR pulse. So, a superposition of electromagnetic pulses emitted at
different moments of time took place within the formation zone of radiation.

In the case of ball, the superposition of electromagnetic pulses emitted at
different moments of time also takes place in the radiation zone. However,
the pulses are emitted by the same particle (not different particles).

\subsection{Amplification of Cherenkov radiation inside a waveguide}

\bigskip \label{section7.2}

Let us consider CR from a chain of equidistant charges moving along the axis
of circular section waveguide filled with transparent and nondispersive
dielectric (e.g.., teflon in the range of wavelengths $\lambda >1mm$ \cite%
{26,27}).

In Fig.\ref{fig4} the first charge is shown to be in point $A$. Let us trace
the propagation of CR pulse $W_{1}\left( A;t\geq t_{A}\right) $ induced by
the first charge and emitted in small vicinity of point $A$ at the instant $%
t_{A}$ (wave line in Fig.\ref{fig4}). At its propagation the pulse $%
W_{1}\left( A;t\geq t_{A}\right) $\ at some instant $t_{B}$\ will cross the
trajectory of charges (axis of waveguide) in the vicinity of some point $%
B^{\ast }$\ at the same angle $\theta $, at which it was emitted. At that
moment $W_{1}\left( A;t\geq t_{A}\right) $\ will fall behind its
faster-than-light source that will be found more to the right at some point $%
B$.

The second charge of the chain is more to the left at the distance of $d$
from the first one and if $d\cong B^{\ast }B$\ it will cross the vicinity of
point $B^{\ast }$\ simultaneously with the pulse $W_{1}\left( A;t\geq
t_{A}\right) $. CR pulse $W_{2}\left( B^{\ast };t\geq t_{B}\right) $
generated by the second charge in the vicinity of point $B^{\ast }$\ will be
formed and simultaneously interfere with $W_{1}\left( A;t\geq t_{A}\right) $
for all $t\geq t_{B}$. If the direction of CR $\cos \theta =c/$v$\sqrt{%
\varepsilon }$, and the radius $a$ of waveguide are known, one can easily
calculate the distance
\begin{equation}
B^{\ast }B=2a\tan \theta  \label{B*B}
\end{equation}%
of a pulse $W_{1}\left( A;t\geq t_{A}\right) $\ from its source at the
instant $t_{B}$. It is clear that at \ $d\cong B^{\ast }B=2a\tan \theta $
the generation of pulses in regions adjacent to the second, third and
subsequent charges will take place simultaneously with their interference
with the pulses emitted earlier by charges propagating ahead of them. Note,
that a similar process may take place for all \textit{d} multiple to the
length of $B^{\ast }B$ and, in particular, at $x\equiv d/B^{\ast }B\cong
1;2;3...$.

The superposition of pulses may be accompanied by their suppression or
amplification depending on this or that value of the phase difference. For
some values of $x\equiv x^{\ast }$ the in-phase superimposition of
oscillations is possible.

Analytical and numerical calculations carried out in \cite{30} have shown
that the in-phase superposition takes place if
\begin{itemize}
    \item $x^{\ast }\cong 1$ (the case of power
quasi-monochromatic CR) and
    \item $x^{\ast }=4$\ (powerful CR in wider frequency band).
\end{itemize}
\noindent In both the cases the pulses are superimposed in-phase
in regions directly adjacent to the radiating charges. But for all
that the superimposed pulses are generated by different charges.
In case of ball the pulses are emitted by the same particle.

\section{Conclusions}

\hspace{0.5cm}1) We have studied the radiation from a relativistic electron
at uniform rotation about a dielectric ball with due regard for dielectric
losses of energy inside the ball material. Here, in addition to the
synchrotron radiation the electron may also emit Cherenkov radiation. Its
rise is attributed to the fact that the field coupled with the relativistic
electron partially penetrates the ball and rotates together with the
particle. At small distances from the surface of ball: $r_{e}\approx r_{b}$,
the speed of this field displacement may be larger than the phase speed of
light in the ball material and then Cherenkov radiation with wavelengths (%
\ref{wavelength}) is emitted inside the ball. The peculiarities in total
radiation at different harmonics $\omega _{k}$ due to the influence of
matter and ball radius are investigated theoretically.

2) In general, the influence of ball on radiation from the particle is small
and radiation differs slightly from synchrotron radiation. However, in case
of weak absorption ($\varepsilon _{b}^{^{\prime \prime }}<<\varepsilon
_{b}^{^{\prime }}$) in the ball material at some harmonics with $k>>1$\ the
electron may generate $n_{k}\geq 1$\ quanta of electromagnetic field during
one rotation period. E.g., an electron of $2MeV$ energy uniformly rotating
at the distance of $0.7mm$ from the surface of fused quartz ball with radius
$r_{b}=3.62cm$ will generate approximately one quantum $\left( n_{8}\approx
1.0\right) $ of electromagnetic field at the wavelength $\lambda _{8}\approx
3cm$\ (8th harmonic) during one revolution (see the highest peak in Fig.\ref%
{fig2}). This value is more than 30-fold greater than the similar value\ of $%
n_{8}$ for electron rotating in a continuous, infinite and transparent
medium having the same real part $\varepsilon ^{^{\prime }}$\ of
permittivity as that for the ball (see (\ref{nkbesk}) and the dashed line in
Fig.\ref{fig2}).

3) Such a magnification of $n_{k}$ is due to the presence of external source
of energy stimulating a uniform rotation of electron around the ball. It is
possible only if
\begin{itemize}
    \item the ratio between the radius of ball and that of
electron orbit takes on a number of fixed values (see, e.g.,
(\ref{n8ball/n8infinity}))
    \item the Cherenkov condition (\ref{Cherenkov}) for the speed of
particle \textquotedblleft image\textquotedblright\ on the ball
surface and the ball material is satisfied and
    \item inside the ball the Cherenkov radiation field passes the
distances that are many times greater than the radius $r_{b}$ of
ball (before the radiation will be absorbed or leave the ball
boundaries).
\end{itemize}

4) High power radiation rises due to the fact that electromagnetic
oscillations of Cherenkov radiation induced by the particle along all its
trajectory are partially locked inside the ball and are superimposed in
nondestructive way (Section \ref{section6}).

\section*{Acknowledgements}

The authors are thankful to Professor A.R. Mkrtchyan for general
encouragement and A.A. Saharian, A.S. Kotanjyan for stimulating discussions.
One of the authors (S.R.A.) is especially thankful to S. Bellucci for
attention and assistance.

The work is supported by Ministry of Education and Science of RA (Grant No.
0063).


\bigskip

\begin{thebibliography}{99}
\bibitem{1} E-E Koch (ed.), Handbook on Synchrotron Radiation, North
Holland, Amsterdam, 1983.

\bibitem{2} I.M. Ternov, V.V. Mikhailin, V.R. Khalilov, Synchrotron
Radiation and Its Applications, Harwood Academic, Amsterdam, 1985.

\bibitem{3} A.A. Sokolov, I.M. Ternov, Radiation from Relativistic Electron,
ATP, New York, 1986.

\bibitem{4} V.A. Bordovitsyn (ed.), Synchrotron Radiation Theory and Its
Development, World Scientific, Singapore, 1999.

\bibitem{5} H. Hiedemann, Synchrotron Radiation, Springer-Verlag, Germany,
2003.

\bibitem{6} A. Hofman, The Physics of Synchrotron Radiation, Cambridge
University Press, Cambridge, 2004.

\bibitem{7} G.B. Rybicky, A.P. Lightman, Radiative Processes in
Astrophysics, J. Wiley, New York, 1979.

\bibitem{8} J.V. Jelley, Cherenkov Radiation and Its Applications, Pergamon
Press, London, 1958.

\bibitem{9} B.M. Bolotovskii, Physics-Uspekhi 75 (1961) 295.

\bibitem{10} V.P. Zrelov, Vavilov-Cherenkov Radiation (and Its Applications
in High Energy Physics), Atomizdat, Moscow, 1968 (in Russian).

\bibitem{11} P. Rullhusen, X. Artru, P. Dhez, Novel Radiation Sources Using
Relativistic Electrons, World Scientific, Singapore, 1998.

\bibitem{12} V.N. Tsytovich, Westnik MGU 11 (1951) 27 (in Russian).

\bibitem{13} K. Kitao, Prog. Theor. Phys. 23 (1960) 759.

\bibitem{14} S.R. Arzumanyan, L.Sh. Grigoryan, A.A. Saharian, Izv. Nats.
Akad. Nauk Arm., Fiz. (Engl. Transl.: J. Contemp. Phys.) 30 (1995) 99.

\bibitem{15} S.R. Arzumanyan, L.Sh. Grigoryan, A.A. Saharian, Kh.V.
Kotanjian, Izv. Nats. Akad. Nauk Arm., Fiz. (Engl. Transl.: J. Contemp.
Phys.) 30 (1995) 106.

\bibitem{16} L.Sh. Grigoryan, A.S. Kotanjyan, A.A. Saharian, Izv. Nats.
Akad. Nauk Arm., Fiz. (Engl. Transl.: J. Contemp. Phys.) 30 (1995) 239.

\bibitem{17} L.Sh. Grigoryan, H.F. Khachatryan, S.R. Arzumanyan, Izv. Nats.
Akad. Nauk Arm., Fiz. (Engl. Transl.: J. Contemp. Phys.) 33 (1998) 267
(Preprint cond-mat/0001322).

\bibitem{18} L.Sh. Grigoryan, H.F. Khachatryan, S.R. Arzumanyan, Izv. Nats.
Akad. Nauk Arm., Fiz. (Engl. Transl.: J. Contemp. Phys.) 37 (2002) 327.

\bibitem{Mher} M.L. Grigoryan, V National Conference on Application of the
X-rays, Synchrotron Radiation, Neutrons, and Electrons for Studying
NANO-Materials (RSNE NANO-2005), 14-19 November 2005, Moscow, Book of
Abstracts, p.302 (in Russian); preprint hep-th/0512080.

\bibitem{19} A.S. Kotanjyan, A.A. Saharian, Izv. Nats. Akad. Nauk Arm., Fiz.
(Engl. Transl.: J. Contemp. Phys.) 37 (2002) 135.

\bibitem{20} A.A. Saharian, A.S. Kotanjyan, Izv. Nats. Akad. Nauk Arm., Fiz.
(Engl. Transl.: J. Contemp. Phys.) 38 (2003) 288.

\bibitem{21} A.A. Saharian, A.S. Kotanjyan, Nucl. Instr. and Meth. B 226
(2004) 351.

\bibitem{22} A.S. Kotanjyan, H.F. Khachtaryan, A.V. Petrosyan, A.A.
Saharian, Izv. Nats. Akad. Nauk Arm., Fiz. (Engl. Transl.: J. Contemp.
Phys.) 35 (2000) 115.

\bibitem{23} A.S. Kotanjyan, A.A. Saharian, Izv. Nats. Akad. Nauk Arm., Fiz.
(Engl. Transl.: J. Contemp. Phys.) 36 (2001) 310; ibid 37 (2002) 263; Mod.
Phys. Lett. A 17 (2002) 1323.

\bibitem{24} A.S. Kotanjyan, Nucl. Instr. and Meth. B 201 (2003) 3.

\bibitem{25} A.A. Saharian, A.S. Kotanjyan, J. Phys. A 38 (2005) 4275.

\bibitem{26} G.C. Southworth, Principles and Applications of Waveguide
Transmission, New York, 1950.

\bibitem{27} E.M. Voronkova, B.N. Grechushnkiov, G.I. Distler, I.P. Petrov,
Optical Materials for Infrared Technology, Nauka, Moscow, 1965 (in Russian).

\bibitem{28} H.C. Lihn, D. Bocek, M. Hernandez, P. Kung, C. Settakorn, H.
Wiedemann, Phys. Rev. Lett. 76 (1996) 4163.

\bibitem{29} C. Settakorn, M. Hernandez, H. Wiedemann, Stimulated Transition
Radiation in the Far-Infrared, SLAC-PUB-7587, August 1997.

\bibitem{30} L.Sh. Grigoryan, H.F. Khachatryan, A.A. Saharian, Kh.V.
Kotanjyan, S.R. Arzumanyan, M.L. Grigoryan, Izv. Nats. Akad. Nauk Arm., Fiz.
(Engl. Transl.: J. Contemp. Phys.) 40 (2005) 155.
\end{thebibliography}
\end{document}